\documentclass[conference]{IEEEtran}
\IEEEoverridecommandlockouts
\usepackage{cite}
\usepackage{algorithmic}
\usepackage{graphicx}
\usepackage{textcomp}
\usepackage{xcolor}
\usepackage{newtxtext}
\usepackage{amsthm}
\usepackage{newtxmath}
\usepackage{tikz}
\def\hb{\hbox to 11.5 cm{}}
\usepackage{enumerate}
\DeclareMathOperator*{\argmax}{argmax} 
\theoremstyle{definition}
\newtheorem{definition}{Definition}
\newtheorem{assumption}{Assumption}
\newtheorem{example}{Example}
\newtheorem{theorem}{Theorem}
\usepackage{enumitem}
\newtheorem{proposition}{Proposition}
\theoremstyle{theorem}
\theoremstyle{proposition}
\DeclareMathOperator*{\argmin}{argmin}
\usepackage{subcaption}
\usepackage{graphicx}
\usepackage{setspace}
\def\BibTeX{{\rm B\kern-.05em{\sc i\kern-.025em b}\kern-.08em
    T\kern-.1667em\lower.7ex\hbox{E}\kern-.125emX}}
\begin{document}

\title{\Large Reconciling Explanations in Multi-Model Systems through Probabilistic Argumentation

}
\author{
	\IEEEauthorblockN{Shengxin Hong$^{1}$, Xiuyi Fan$^{2}$}
	\IEEEauthorblockA{$^1$ Hubei University of Technology, Wuhan, China}
	\IEEEauthorblockA{$^2$ Nanyang Technological University, Singapore}
}

\maketitle
\begin{abstract}
Explainable Artificial Intelligence (XAI) has become critical in enhancing the transparency and trustworthiness of AI systems, especially as these systems are increasingly deployed in high-stakes domains such as healthcare and finance. Despite the progress made in developing explanation generation techniques for individual machine learning (ML) models, significant challenges remain in achieving coherent and comprehensive explanations in multi-model systems. This paper addresses these challenges by focusing on the explanation reconciliation problem (ERP) within multi-model systems. Traditional explanation generation technique often fall short in multi-model systems contexts, where explanations from different models can conflict and fail to form a cohesive narrative. Through the use of probabilistic argumentation and knowledge representation techniques, we propose a framework for generating holistic explanations that align with human cognitive processes. Our approach involves mapping uncertain explanation information to probabilistic arguments and introducing criteria for explanation reconciliation based on user perspectives such as optimism, pessimism, fairness. In addition, we introduce the relative independence assumption to optimise the search space for computational explanations.
\end{abstract}


\section{Introduction}

Explainable Artificial Intelligence (XAI) has emerged as a pivotal area of research within the broader field of AI, driven by the increasing complexity and opacity of modern machine learning (ML) models \cite{gunning2019xai}. As AI systems become more pervasive in critical applications such as healthcare, finance, and autonomous systems \cite{hong2024argmed}, the need for transparency, interpretability, and trustworthiness has never been more pronounced. These qualities are key to understanding, diagnosing, and mitigating biases and errors in automated decision-making systems. XAI bridges this transparency and interpretability gap by developing methods and tools that make AI decisions comprehensible to humans, enabling users to trust, manage, and govern these systems responsibly.

In this context, significant efforts have been dedicated to enhancing the explainability of ML models. Researchers have investigated various strategies to make individual ML models more comprehensible to users. Common approaches include Local Interpretable Model-agnostic Explanations (LIME)\cite{9643257}, SHapley Additive exPlanations (SHAP) \cite{antwarg2021explaining}, Counterfactual Explanations (CE) \cite{guidotti2022counterfactual}, and Predictive Uncertainty Estimation (PUE) \cite{lakshminarayanan2017simple}. By demonstrating how input features affect prediction outcomes, these techniques enhance users' comprehension of individual ML models, which are designed to produce direct predictions from given inputs.

One critical challenge remains. In many real-world applications, multiple sources of information are used, and chains of decisions must be made. Thus, explaining relations between directly connected inputs and outputs becomes insufficient for understanding the complex system holistically. In other words, complex AI systems can face the following explanation challenges:
\begin{itemize}[leftmargin=*]
	\item[](1) A complete AI system contains multiple components and the input of one component may be the output of another. It is difficult to aggregate all the local explanations into a coherent narrative that accurately reflects the collective behavior of the multi-component system.
	\item[](2) Since incomplete information and AI agents operate in different probability spaces, explanations from different models may conflict with each other. Thus, means for reconciliating conflicting information must be developed.
\end{itemize}

\noindent We illustrate this multi-component system with a clinical diagnostic example, as shown in Example \ref{E1}:
\begin{example}
	\label{E1}
	Consider a brain disease diagnostic system that contains four functionally different ML models of $a,b,c,d$ that are used for brain MRI image analysis, cognitive test analysis, brain risk assessment, and diagnosis, respectively.
	\vspace{-0.6cm}
	\begin{figure}[htbp]
		\centering\includegraphics[width=0.44\textwidth,height=0.21\textwidth]{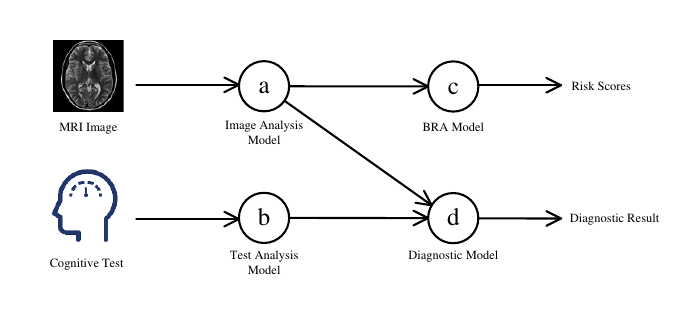} 
		\vspace{-0.65cm}
	\end{figure}

	\noindent From these components, we can obtain detailed probability information for each. For instance: (1) For an input MRI image, model $a$ predicts a brain abnormality with a probability of 0.7, whereas the cognitive test model $b$ predicts a cognitive abnormality with a probability of 0.4. (2) Assuming that the patient has a brain abnormality, the probability of model $d$ diagnosing them with Alzheimer’s disease is 0.5. However, reconciling these probabilities into a coherent explanation is not trivial, as individual model probabilities may conflict with each other. For example, model $b$ shows a higher probability that the cognitive test results in a cognitive abnormality, which usually indicates early Alzheimer’s disease. However, model $c$ shows a lower risk of brain disease, which is not consistent with results of model $b$.
\end{example}

From Example 1, we can see that although it is critical to apply traditional XAI techniques to explain how the MRI Analysis model derives its 0.7 prediction by, for instance, highlighting parts of the image that lead to this prediction, it is equally important to untangle the relations between different components and derive a coherent understanding of the entire system. To this end, we consider a version of the Explanation Reconciliation Problem (ERP) introduced in \cite{vasileiou2024generating}.

Roughly speaking, consider a complex system in which each model has an associated probability. There are probabilistic dependencies among models, and the probabilities of the models may or may not be independent. We aim to develop a framework in which all these probabilities can be captured and correctly processed.

To this end, we develop a probabilistic argumentation-based framework with probabilistic arguments and rules. We represent prediction models as probabilistic arguments and study their probabilistic relations. We propose three distinct perspectives on handling uncertainty, with the goal of finding a set of holistic explanations in a multi-model system.

The rest of this paper is organised as the follows. Section 2 reviews concepts introduced in the literature that are used in this work. Section 3 introduce the formal representations of explanations of models and how to map explanations to probabilistic rules and the concept of holistic explanations. In Section 4, we introduce the relative independence assumption to optimise the search space for computational explanations. Section 5 discusses our work in comparison with related work. We conclude in Section 6.
\section{Background and Preliminaries}
\label{s2}
Probabilistic Rules (p-rules) \cite{fan2022probabilistic} serve as the primary building blocks for probabilistic structured argumentation. They enable the construction of arguments that accommodate uncertainty and offer a conditional probability interpretation to rules commonly employed in structured argumentation.

\begin{definition}\cite{fan2022rule} 
	Given a language $\mathcal{L}$, a \textit{probabilistic rule (p-rule)} is $\sigma_0\leftarrow\sigma_1,...,\sigma_k:[\theta]$, for $k\geq0,\sigma_i\in\mathcal{L},0\leq\theta\leq 1$.
\end{definition}

\begin{definition}\cite{HENDERSON2020103199}
	Given a language $\mathcal{L}$ with $n$ sentences, the \textit{Complete Conjunction Set (CC Set)} $\Omega$ of $\mathcal{L}$ is the set of $2^n$ conjunction of sentences such that each conjunction	containsc $n$ distinct sentences.
\end{definition}

\begin{definition}\cite{fan2022rule}
	Given a language $\mathcal{L}$ and p-rules $\mathcal{R}$, $\Omega$ is the CC set of $\mathcal{L}$. A function $\pi:\Omega\rightarrow[0,1]$ is a consistent probability distribution with respect to $\mathcal{R}$ on $\mathcal{L}$ for $\Omega$ iff:
	\begin{enumerate}
		\item For all $\omega_i\in\Omega$, $0\leq\pi(\omega_i)\leq 1$, it holds that:
		\begin{equation}
			\sum_{\omega_i\in\Omega}\pi(\omega_i)=1
		\end{equation}
		\item For each p-rule $\sigma_0\leftarrow:[\theta]\in\mathcal{R}$, it holds that:
		\begin{equation}
			\label{e1}
			\theta=\sum_{\omega_i\in\Omega,\omega_i\models\sigma_0}\pi(\omega_i)
		\end{equation}
		\item  For each p-rule $\sigma_0\leftarrow\sigma_1,...,\sigma_k:[\theta]\in\mathcal{R},(k>0)$, it holds that:
		\begin{equation}
			\label{e3}
			\setlength{\belowdisplayskip}{5pt}
			\sum_{\omega_i\in\Omega,\omega_i\models\sigma_1\wedge,...,\wedge\sigma_k}\pi(\omega_i)\times\theta=\sum_{\omega_i\in\Omega,\omega_i\models\sigma_0\wedge,...,\wedge\sigma_k}\pi(\omega_i)
		\end{equation}
	\end{enumerate}
	\label{d2.5}
\end{definition}\label{d2.6}
Given a language $\mathcal{L}$ and a set of p-rule $\mathcal{R}$, let $\Pi$ be the set of consistent probability distributions wrt $\mathcal{R}$ on $\mathcal{L}$. There are three kinds of reasoning asserts for $\sigma\in\mathcal{L}$:
\begin{itemize}[noitemsep,nolistsep]
	\item \emph{$\sigma$-maixmal Solution}, a upper bounds distribution $\pi_0\in\Pi$ of $\Pr(\sigma)$ as follows.
	\begin{equation}
		\pi_0=\argmax_{\pi\in \Pi}\sum_{\omega_i\in\Omega,\omega_i\models\sigma}\pi(\omega_i)
	\end{equation}
	\item \emph{$\sigma$-minimal Solution}, a lower bounds distribution $\pi_0\in\Pi$ of $Pr(\sigma_0)$ as follows.
	\begin{equation}
		\pi_0=\argmin_{\pi\in \Pi}\sum_{\omega_i\in\Omega,\omega_i\models\sigma}\pi(\omega_i)
	\end{equation}
	\item \emph{Maximum Entropy Solution,} a Maximum Entropy Distribution $\pi_0\in\Pi$ is:
	\begin{equation}
		\setlength{\belowdisplayskip}{5pt}
		\pi_0=\argmax_{\pi\in\Pi}(-\sum_{\omega_i\in\Omega}\pi(\omega_i)\log(\pi(\omega_i)))
	\end{equation}
\end{itemize}
As explained in \cite{fan2022rule}, solving linear systems derived from p-rules to compute the joint distribution $\pi$ may result in multiple solutions, as the linear system could be underdetermined. Thus, three reasoning paradigms are proposed to represent three potential selections of distributions. $\sigma$-maximal and $\sigma$-minimal solutions aim to maximize and minimize the probability of a chosen $\sigma$, respectively, while the maximum entropy solution aims to minimize selection bias in choosing solutions.

\section{Explanation Reconciliation}
\label{padf}
In this section, we first describe how to formally represent prediction models.
\begin{definition}
	\label{d1}
	A \emph{model} is a tuple $\mathcal{M} = (\mathcal{I},\mathcal{O},\mu)$ such that:
	\begin{itemize}
		\item $\mathcal{I} = \mathcal{I}_e \cup \mathcal{I}_w$ is a set of input of the model such that $\mathcal{I}_e$ is the external input and $\mathcal{I}_w$ is the internal input;
		\item $\mathcal{O}$ is a set of output of the model;
		\item $\mu:\mathcal{I}\times\mathcal{O}\rightarrow [0,1]$ is a mapping from input to output.
	\end{itemize}
	 
\end{definition}
In Definition \ref{d1}, external inputs are known conditions that are inputs from outside the system, while internal inputs originate from other models output within the system. $\mu$ is a mapping function for generating explanations which can map the probabilities of model's different inputs to outputs, and this mapping explains the dependencies between different features and output.

For a system consisting with multiple prediction models, it has multiple explanations derived from different models. These interconnected form models can be formalised as follows:
\begin{definition}
\label{d5}
	A \emph{multi-model system} is a graph $\mathcal{S}=(\{\mathcal{M}\}_n, V)$ and $n\geq 1$ such that:
	\begin{itemize}
		\item $V$ is a set of \emph{directed link} in models such that if $(\mathcal{M}_i,\mathcal{M}_k)\in V, i\neq k$ and $1\leq i,k\leq n$, then the internal input of $\mathcal{M}_k$ is the output of $\mathcal{M}_i$ denoted as $\mathcal{I}^k_w = \mathcal{O}_i$.
		\item If there not exist any $i,k\in n,(\mathcal{M}_i,\mathcal{M}_k)\in V$, then $\mathcal{O}_i$ is one of the \emph{final output} of this multi-model system.
	\end{itemize}
\end{definition}
Definition \ref{d5} provides a formal representation of multiple models. We illustrate this definition with the following example.
\begin{example}
	\label{e2}
	(Example \ref{E1} continued) For the brain disease diagnostic system in Example \ref{E1}, it have following models:
	\begin{itemize}
		\item $\mathcal{M}_a:$ Given an input MRI image, the probability of the analysis report showing a Brain abnormality (BA) is 0.7;
		\item $\mathcal{M}_b:$ Given an input cognitive test (CT), the probability of the analysis report showing a cognitive abnormality (CA) is 0.6;
		\item $\mathcal{M}_c:$ Assuming that the patient's analysis report shows a brain abnormality, model $c$ has a probability of 0.2 to output that the patient has a high risk (HR) of brain disease;
		\item $\mathcal{M}_d:$ Assuming that the patient's analysis report shows a brain abnormality, model $d$ has a probability of 0.6 to 
		diagnose patients with early Alzheimer's disease (AD);
		\item $\mathcal{M}_d:$ Assuming that the patient's analysis report shows a cognitive abnormality, model $d$ has a probability of 0.5 to 
		diagnose patients with early Alzheimer's disease (AD);
	\end{itemize}
Note that there are more inputs and outputs for each models. To reduce the complexity of the example, only these five inputs and outputs are presented. Hence, $\mathcal{S}$ can be formalised as:

\noindent$\{\mathcal{M}\}_n$ consists of:
\begin{itemize}
	\item[] $\mathcal{M}_a=(\{MRI\}_{\mathcal{I}_e}\cup\{\emptyset\}_{\mathcal{I}_w},\{BA\}_{\mathcal{O}})$
	\item[] $\mathcal{M}_b=(\{CT\}_{\mathcal{I}_e}\cup\{\emptyset\}_{\mathcal{I}_w},\{CA\}_{\mathcal{O}})$
	\item[] $\mathcal{M}_c=(\{\emptyset\}_{\mathcal{I}_e}\cup\{BA\}_{\mathcal{I}_w},\{HR\}_{\mathcal{O}})$
	\item[] $\mathcal{M}_d=(\{\emptyset\}_{\mathcal{I}_e}\cup\{BA,CA\}_{\mathcal{I}_w},\{AD\}_{\mathcal{O}})$
\end{itemize}
$V$ consists of:
\begin{itemize}
	\item[] ($\mathcal{M}_a,\mathcal{M}_c$), \space($\mathcal{M}_a,\mathcal{M}_d$), \space($\mathcal{M}_b,\mathcal{M}_d$)
\end{itemize}
$\mu$ consists of:
\begin{itemize}
	\item[] $\mu(BA|MRI)=0.7,$\space\space\space$\mu(HR|BA)=0.2$\space\space\space$\mu(AD|BA)=0.6$
	\item[] $\mu(CA|CT)=0.6,$\space\space\space\space\space$\mu(AD|CA)=0.5$
\end{itemize}
\renewcommand{\thefigure}{2}
In addition, it is easy to see that the final output is $HR$ and $AD$. $\mathcal{S}$ is illustrated in Figure 2.
\vspace{-0.5cm}
\begin{figure}[htbp]
	\centering\includegraphics[width=0.44\textwidth,height=0.16\textwidth]{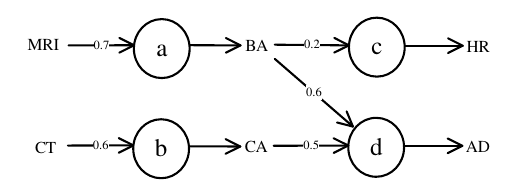} 
	\vspace{-0.2cm}
	\caption{Illustration of Example 2. The nodes represent the model, the pointing arrows represent inputs and outputs of models.}
\end{figure}
\end{example}
\vspace{-0.4cm}
With the multi-models systems defined, we introduce probabilistic argumentation and define the \emph{holistic explanation} of multi-model systems, formally:
\begin{definition}
	\label{d6}
	Given a multi-model system $\mathcal{S}=(\{\mathcal{M}_n\},V)$, it have corresponding language $\mathcal{L}$ and a set of p-rules $\mathcal{R}$ such that:
	\begin{itemize}
		\item $\mathcal{L}=\{\sigma_i\vert \sigma_i\in \mathcal{O}_k, k\in n\}$;
		\item $\mathcal{R}$ consists of:
		
		$\{\sigma_0\leftarrow :[\theta]| \theta =\mu(\sigma_0|\sigma_1,...,\sigma_k), \sigma_i\in\mathcal{I}_e,1\leq i\leq k\}$
		                                                            
		$\{\sigma_0\leftarrow \sigma_1,...,\sigma_k:[\theta]| \theta =\mu(\sigma_0|\sigma_1,...,\sigma_k),\sigma_i\in\mathcal{I}_w\}$
	\end{itemize}
\end{definition}
Since the external inputs are always known facts, we do not need to consider the external inputs when mapping $\mathcal{L}$ which are defined as empty in $\mathcal{R}$. We will illustrate how to map PADF to a language $\mathcal{L}$ and a set of p-rules $\mathcal{R}$ with the following example.

\begin{example}
	\label{e4}
	(Example \ref{e2} continued) Given a multi-model system $\mathcal{S}=(\{\mathcal{M}_n\},V)$ in Example \ref{E1}, its corresponding language $\mathcal{L}$ and a set of p-rules $\mathcal{R}$ are:
	\begin{itemize}
		\item $\mathcal{L} = \{BA,CA,HR,AD\}$
		\item $\mathcal{R}$ consists of:
		 \item[] $BA\leftarrow:[0.7],$ \space\space $HR\leftarrow BA:[0.2],$\space\space $AD\leftarrow BA:[0.6],$
		 \item[] $CA\leftarrow:[0.6],$ \space\space $AD\leftarrow CA:[0.5]$
	\end{itemize}
\end{example}
By reconciling multiple models, we define the holistic explanation obtained by reconciling explanations from multiple models as the consistent probability distribution of the explanandum, formally:
\begin{definition}
	Given a multi-model system $\mathcal{S}=(\{\mathcal{M}_n\},V)$ with its corresponding language $\mathcal{L}$ and a set of p-rules $\mathcal{R}$. $\phi\in\mathcal{L}$ is an  explanandum, it have \emph{holistic explanation} $e$ if and only if the following two conditions holds:
	\begin{itemize}[leftmargin=*,noitemsep,nolistsep]
		\item[] C1. $\phi$ is a \emph{final output} in $\mathcal{S}$;
		\item[] C2. Let $E$ be the set of consistent probability distributions wrt related factors of $\phi$, then $e\in E$.
	\end{itemize}
\label{d7}
\end{definition}

In Definition \ref{d7}, the  related factors of $\phi$ are all the $\sigma\in\mathcal{L}$ which is related to $\phi$.

However, when the informations of models are incomplete or conflicting, the consistent probability distribution wrt related factors of $\phi$ will have an infinite number of solutions. To this end, we define three holistic explanation criteria that describe the different perceptions of users about uncertainty explanation information in order to select holistic explanations that are aligned with human cognition:

\begin{definition}
	\label{d3.4}
	\textbf{(Optimistic Criterion)} Given a of multi-model system $\mathcal{S}=(\{\mathcal{M}_n\},V)$ with its corresponding language $\mathcal{L}$ and a set of p-rules $\mathcal{R}$. The \emph{holistic explanation} in $\mathcal{S}$ with the \emph{Optimistic Criterion} is to maximize the probability of $\phi$:
		\begin{equation}
			e_0= \argmax_{e\in E} \Pr(\phi)
			\label{e11}
		\end{equation}
	The Optimistic Criterion describes that users always have optimistic perspectives on uncertainty explanation information. Under Optimisyic Criterion, the user are radical in probabilistic reasoning and tries to maximise the prediction result.
\end{definition}

\begin{definition}
	\label{d3.5}
	\textbf{(Pessimistic Criterion)} Given a of multi-model system $\mathcal{S}=(\{\mathcal{M}_n\},V)$ with its corresponding language $\mathcal{L}$ and a set of p-rules $\mathcal{R}$. The \emph{holistic explanation} in $\mathcal{S}$ with the \emph{Pessimistic Criterion} is to minimize the probability of $\phi$:
	\begin{equation}
		e_0= \argmin_{e\in E} \Pr(\phi)
		\label{e13}
	\end{equation}
The opposite of the Optimistic Criterion is the Pessimistic Criterion. In this Criterion, users are always conservative in probabilistic reasoning and tring to minimise the prediction result.
\end{definition}

\begin{definition}
	\label{d3.7}
	\textbf{(Laplace Criterion)} Given a multi-model system $\mathcal{S}=(\{\mathcal{M}_n\},V)$ with its corresponding language $\mathcal{L}$ and a set of p-rules $\mathcal{R}$.  The \emph{holistic explanation} in $\mathcal{S}$ with the \emph{Laplace Criterion} defines as follows:
	\begin{equation}
		e_0=\argmax_{e\in E}(-\sum_{e_i\in E}\pi(e_i)\log(\pi(e_i)))
		\label{e15}
	\end{equation}
\end{definition}
Laplace Criterion assumes that the absence of information means that all outcomes have equal probability. When there are unknown probabilities, a unique set of consistent probability distributions can be determined using the principle of maximum entropy, and it characterises the known information well and distributes the probabilities fairly \cite{PhysRev.106.620}.

\section{Computing Holistic Explanation}
\label{pd}
%The study from \cite{fan2022probabilistic} points out that the time to solve Rule-PSAT increases exponentially as the size of $\mathcal{L}$ grows. This is not conducive to solving for the holistic explanation, as there multiple model explanations can make the search space extraordinarily large. The study \cite{yoon1995multiple} proposes that although there are various links between features, they are often considered independently in decision. On this basis, the work by Hadoux \cite{hadoux2017strategic} proposed a multi-agent persuasion framework based on decision trees and optimises the computation of decision rules through independence assumption. However, while this assumption bypasses the computation of the joint probability distribution, it makes its criterion constrain only the neighbouring nodes of each computational step. In order to solve for the holistic explanation for system, we expect the explanation criteria to constrain all models in a global perspective. To this end, we introduced the relative independence assumption (RIA). First, we define the notion of \emph{reachable}.

From Definition \ref{d7}, we see that to compute a holistic explanation for a multi-model system, we must determine a consistent probability distribution. This is a known NP problem, with complexity increasing exponentially as the number of prediction models grows \cite{fan2022probabilistic}. To ease the computation, one simplification we can impose is assuming independence between models. To this end, we introduce the Relative Independence Assumption (RIA), along with the notion of {\em reachability}, defined as follows.

\begin{definition}
	\label{reach}
Given a multi-model system $\mathcal{S}=(\{\mathcal{M}_n\},V)$ with its corresponding language $\mathcal{L}$ and a set of p-rules $\mathcal{R}$. For any $\sigma_i,\sigma_j\in \mathcal{L}$, we say that $\sigma_i$ is \emph{reachable} from $\sigma_j$ if and only if the following two conditions holds:
	\begin{itemize}[leftmargin=*,noitemsep,nolistsep]
		\item[] C1. there exists a p-rule $\sigma_i\leftarrow \sigma_j,...,\sigma_k:[\theta]$ or
		\item[] C2.  there exists $\sigma'\in \mathcal{L}$ such that $\sigma_i$ is \emph{reachable} from $\sigma'$ and $\sigma'$ is \emph{reachable} from $\sigma_j$.
	\end{itemize}
\end{definition}
Definition \ref{reach} is given recursively with C1 being the base case. With this, RIA on $\mathcal{L}$ can be explained as follows.
\begin{assumption}
	\label{d4.2}\textbf{(Relative Independence Assumption)}
	Given a multi-model system $\mathcal{S}=(\{\mathcal{M}_n\},V)$ with its corresponding language $\mathcal{L}$ and a set of p-rules $\mathcal{R}$. For all $\sigma_i,\sigma_j\in\mathcal{L}$ are \emph{independent} if and only if $\sigma_i$ is not \emph{reachable} from $\sigma_j$ and $\sigma_j$ is not \emph{reachable} from $\sigma_i$.
\end{assumption}
\begin{proposition}
	\label{p2}
	For any explanandum $\phi_i,\phi_k\in\mathcal{L}$ in $\mathcal{S}$, $\phi_i$ and $\phi_k$ are \emph{independent} with each other.
\end{proposition}
The proof of Proposition \ref{p2} comes from Definition \ref{d5} and \ref{d7}, wherec each explanandum $\phi$ is final output that represents no model that receives it and no further successor nodes.
\begin{definition}
		Given a multi-model system $\mathcal{S}=(\{\mathcal{M}_n\},V)$ with its corresponding language $\mathcal{L}$ and a set of p-rules $\mathcal{R}$. For any explanandum $\phi\in \mathcal{L}$, there is a corresponding \emph{reachable set} $\mathcal{L}_{\phi}\subseteq\mathcal{L}$ if and only if for any $\sigma_i\in\mathcal{L}_{\phi}$, $\sigma_i=\phi$ or the $\phi$ is reachable from $\sigma_i$. 
		\label{d12}
\end{definition}
\noindent For $\mathcal{L}_{\phi}$, its p-rules denoted as $\mathcal{R}_{\phi}$. 
\begin{proposition}
	\label{con}
	Given a \emph{reachable set} $\mathcal{L}_{\phi}\subseteq\mathcal{L}$. If and only if $\sigma_i\in \mathcal{L}_{\phi}$, it is the \emph{realted factors} of $\phi$.
\end{proposition}
In Definition \ref{d7} we mentioned that the related factors are all the factors (including itself) that may affect the explanandum. Proposition \ref{con} follows from the RIA derivation that all factors not in \emph{reachable set} $\mathcal{L}_{\phi}\subseteq\mathcal{L}$ are unrelated to $\phi$ because they are mutually unreachable.

In the following theorem, we present a semantic relation between the explanation criteria and p-rules:
\begin{theorem}
	Given a multi-model system $\mathcal{S}=(\{\mathcal{M}_n\},V)$ with its corresponding language $\mathcal{L}$ and a set of p-rules $\mathcal{R}$. For an explanandum $\phi$ in $\mathcal{S}$, $\pi:\Omega\rightarrow[0,1]$ is a consistent probability distribution with respect to $\mathcal{R}_{\phi}$ on $\mathcal{L}_{\phi}$. Let $\psi=oc,pc,lc$ denote optimistic criterion, pessimistic criterion and laplace criterion respectively. The holistic explanations $e_\psi$ for $\phi$ under different explanations criteria is:
	\begin{equation}
		e_\psi=\begin{cases}
			\argmax\limits_{\pi\in \Pi}\sum\limits_{\omega_i\in\Omega,\omega_i\models\phi}\pi(\omega_i),  & \text{if }\psi=oc \\\\
			\argmin\limits_{\pi\in \Pi}\sum\limits_{\omega_i\in\Omega,\omega_i\models\phi}\pi(\omega_i), & \text{if }\psi=pc   \\\\
			\argmax\limits_{\pi\in\Pi}(-\sum\limits_{\omega_i\in\Omega}\pi(\omega_i)\log(\pi(\omega_i))),  & \text{if }\psi=lc
		\end{cases}
	\end{equation}
\label{t1}
\end{theorem}
\vspace{-0.4cm}
\begin{proof} (Sketch.)
	The proof of Theorem \ref{t1} follows from Definition \ref{d7}, Definition \ref{d12}, and Proposition \ref{con}. Definition \ref{d7} states that the holistic explanation of explanandum $\phi\in\mathcal{L}$ is the distribution of related factors. Based on RIA, any factor $\sigma_i\in\mathcal{L}$ outside the reachable set $\mathcal{L}_\phi$ is not related to $\phi$. Thus it can be shown that the holistic explanation of $\phi$ is the consistent distribution of the reachable set.
\end{proof}
\vspace{-0.2cm}
\begin{example}
	(Example \ref{e4} continued) Given a multi-model system $\mathcal{S}=(\{\mathcal{M}_n\},V)$ in Example \ref{E1}, with its corresponding language $\mathcal{L}$ and a set of p-rules $\mathcal{R}$. Now we want to solve the holistic explanations $e_\psi$ for explanandum $\phi = AD$ under different explanations criteria. Its corresponding $\mathcal{R}_\phi$ on reachable set $\mathcal{L}_\phi$ consist of: 
\vspace{0.05cm}

\noindent{\small$BA\leftarrow:[0.7],$ \space $AD\leftarrow BA:[0.6],$\space $CA\leftarrow:[0.6],$ \space $AD\leftarrow CA:[0.5]$}
\vspace{0.05cm}

\noindent The $\mathcal{R}_\phi$ set up its equations as follows by Definition \ref{d2.5}:
\begin{itemize}[leftmargin=*,noitemsep,nolistsep]
	\item[] $(\pi(010)+\pi(011)+\pi(110)+\pi(111))\times0.6=\pi(111)+\pi(110),$
	\item[] $(\pi(001)+\pi(011)+\pi(101)+\pi(111))\times0.5=\pi(111)+\pi(101),$
	\item[] $0.7=\pi(010)+\pi(011)+\pi(110)+\pi(111),$
	\item[] $0.6=\pi(001)+\pi(011)+\pi(101)+\pi(111),$
	\item[] $1=\pi(000)+\pi(001)+\pi(010)+\pi(011)+\pi(100)+\pi(101)+\pi(110)+\pi(111).$
\end{itemize}
Please note that in here, $111$ and $001$ denote that $AD\land CA \land BA$ and $\lnot AD\land \lnot CA \land BA$ respectively.
\vspace{0.05cm}

\noindent Solve these, the holistic explanations $e_0$ under Optimistic Criterion is:
\begin{itemize}[leftmargin=*,noitemsep,nolistsep]
	\item[] {\small$\pi(000)=0,$ \space\space\space\space\space\space\space $\pi(001)=0.02,$\space\space $\pi(010)=0,$\space\space\space\space\space\space\space $\pi(011)=0.28,$}
	\item[] {\small$\pi(100)=0.15,$ \space\space $\pi(101)=0.13,$\space\space $\pi(110)=0.25,$\space\space $\pi(111)=0.17,$}
\end{itemize}
\noindent Solve these, the holistic explanations $e_0$ under Pessimistic Criterion is:
\begin{itemize}[leftmargin=*,noitemsep,nolistsep]
	\item[] {\small$\pi(000)=0.14,$ \space\space $\pi(001)=0.16,$\space\space $\pi(010)=0.14,$\space\space $\pi(011)=0.14,$}
	\item[] {\small$\pi(100)=0,$ \space\space\space\space\space\space\space $\pi(101)=0,$\space\space\space\space\space\space\space $\pi(110)=0.12,$\space\space $\pi(111)=0.3,$}
\end{itemize}
\noindent Solve these, the holistic explanations $e_0$ under Laplace Criterion:
\begin{itemize}[leftmargin=*,noitemsep,nolistsep]
	\item[] {\small$\pi(000)=0.058,$ \space $\pi(001)=0.114,$\space $\pi(010)=0.094,$\space $\pi(011)=0.186,$}
	\item[] {\small$\pi(100)=0.058,$ \space $\pi(101)=0.07,$\space $\pi(110)=0.19,$\space $\pi(111)=0.23,$}
\end{itemize}
With these computing results of holistic explanations, we demonstrate the use of holistic explanations with two cases:
\begin{itemize}
	\item John is a patient. Given John's MRI image and conitive test result (external input), the $\pi(101)$ in holistic explanations for diagnostic result (explanandum) under Laplace Criterion shows that the probability of the patient being diagnosed with early Alzheimer's disease with brain abnormalities and normal cognition is 0.07.
	\item For Sarah's MRI image and conitive test result (external input), the $\pi(011)$ in holistic explanations for diagnostic result (explanandum) under Optimistic Criterion shows that the probability of Sarah has cognitive abnormalities and brain abnormalities but no diagnosis of early Alzheimer's disease is 0.28.
\end{itemize}
\end{example}

\section{Related Works}
As a novel area in Explainable AI, there is very little work in the literature that focuses on explaining the entire system behavior in multi-model systems. As pointed out in \cite{liang2023foundationstrendsmultimodalmachine,engelmann2024mixedmodelsmultipleinstance}.

In \cite{liang2023foundationstrendsmultimodalmachine}, the authors have discussed representation, inference, and explanation in a multi-modal system, which potentiality can include many models. Our work differs from theirs by presenting a formal framework that explicitly compute probability relations between models.

In \cite{engelmann2024mixedmodelsmultipleinstance}, the authors delve into the integration of Generalized Linear Mixed Models (GLMM) with Multiple Instance Learning (MIL) to address the challenge of modeling cell state heterogeneity in single-cell data. The design of MixMIL explains the entire system's behaviour by modelling the dependencies of the different models, our work differs from theirs as our approach further integrates the dependencies between these models through probabilistic argumentation, providing a consistent and coherent explanation in the form of probability distributions.

In probabilistic argumentation, the work by Hadoux \cite{hadoux2017strategic} proposed a multi-agent persuasion framework based on decision trees and optimises the computation of decision rules through independence assumption. However, while this assumption bypasses the computation of the joint probability distribution, it makes its criterion constrain only the neighbouring nodes of each computational step. In contrast, our approach preserves the global nature of the holistic explanation by introducing the relative independence assumption (RIA).

In the field of Probabilistic Logic Programming (PLP) \cite{Fierens2011InferenceIP}, the authors also explored the concept of holistic explanation. They associate holistic explanations with possible worlds, which is consistent with our view. For example, the MPE (Most Probable Explanation) task consists of finding the world with the highest probability given some evidence \cite{84b457d56ae845d883f85eff5cd0b259}. Our work differs from theirs as we use the consistent distribution as an explanation and we provide users with three probabilistic perspectives on the explanation criterion.

\section{Conclusion}
\label{s6}
Explainable AI has been in rapid development over the past decade, with many successful algorithms developed for explaining standalone machine learning models. With these advancements, people can draw insights on the relationships between input features and individual prediction outcomes. However, there is little work on helping users understand complex systems composed of multiple connected machine learning models. These models can be connected to each other or share inputs. They make related but distinct predictions and contribute to overall AI decision-making. 

In this work, we present the Model Reconciliation solution, which formally models such complex AI systems using probabilistic argumentation. Supported by rigorous computation, our explanation answers questions such as, “What is the probability of obtaining output X, when sub-model I predicts A and sub-model II predicts B?” Such detailed understanding of the underlying probabilistic space of the multi-model system provides in-depth insights into the system.

This work presents the theoretical underpinnings of such probabilistic modeling, as well as computational techniques for efficiently computing such explanations. Looking ahead, we aim to explore applications of our framework across various domains, thereby further enhancing the transparency and trustworthiness of AI-driven decision-making.

\vspace{-0.2cm}

\bibliographystyle{IEEEtran}
\bibliography{IEEEfull}
\end{document}